\documentclass[aps,pre,reprint,superscriptaddress,showpacs]{revtex4-1}

\usepackage{graphicx}
\def\micron{\hbox{$\mu$m}}
\usepackage{color}
\newcommand{\Q}[1]{{{\textcolor{black}{#1}}}}
\newcommand{\QQ}[1]{{{\textcolor{black}{#1}}}}
\newcommand{\QQQ}[1]{{{\textcolor{black}{#1}}}}

\begin{document}


\title{Particle dynamics in two-dimensional random energy landscapes -- experiments and simulations}


\author{Florian Evers}
\email[]{florian.evers@hhu.de}
\affiliation{Condensed Matter Physics Laboratory, Heinrich-Heine-University, Universit\"atsstra{\ss}e 1, 40225 D\"usseldorf, Germany}

\author{Christoph Zunke}
\affiliation{Condensed Matter Physics Laboratory, Heinrich-Heine-University, Universit\"atsstra{\ss}e 1, 40225 D\"usseldorf, Germany}

\author{Richard D.~L.~Hanes}
\affiliation{Condensed Matter Physics Laboratory, Heinrich-Heine-University, Universit\"atsstra{\ss}e 1, 40225 D\"usseldorf, Germany}

\author{J\"{o}rg Bewerunge}
\affiliation{Condensed Matter Physics Laboratory, Heinrich-Heine-University, Universit\"atsstra{\ss}e 1, 40225 D\"usseldorf, Germany}

\author{Imad Ladadwa}
\affiliation{Institut f\"ur Physikalische Chemie, Universit\"at M\"unster, 48149 M\"unster, Germany}
\affiliation{Fahad Bin Sultan University, 71454 Tabuk, Saudi-Arabia}

\author{Andreas Heuer}
\affiliation{Institut f\"ur Physikalische Chemie, Universit\"at M\"unster, 48149 M\"unster, Germany}

\author{Stefan U.~Egelhaaf}
\affiliation{Condensed Matter Physics Laboratory, Heinrich-Heine-University, Universit\"atsstra{\ss}e 1, 40225 D\"usseldorf, Germany}


\date{\today}

\begin{abstract}
The dynamics of individual colloidal particles in random potential energy landscapes were investigated experimentally and by Monte Carlo simulations. The value of the potential at each point in the two-dimensional energy landscape follows a Gaussian distribution. The width of the distribution, and hence the degree of roughness of the energy landscape, was varied and its effect on the particle dynamics studied. \QQ{This situation represents an example of Brownian dynamics in the presence of disorder.} In the experiments, the energy landscapes were generated optically using a holographic set-up with a spatial light modulator, and the particle trajectories were followed by video microscopy. The dynamics are characterized using, e.g., the time-dependent diffusion coefficient, the mean squared displacement, the van Hove function and the non-Gaussian parameter. In both, experiments and simulations, the dynamics are initially diffusive, show an extended sub-diffusive regime at intermediate times before diffusive motion is recovered at very long times. \QQ{The dependence of the long-time diffusion coefficient on the width of the Gaussian distribution agrees with theoretical predictions.} Compared to the dynamics in a one-dimensional potential energy landscape, the localization at intermediate times is weaker and the diffusive regime at long times reached earlier, which is due to the possibility to avoid local maxima in two-dimensional energy landscapes.\end{abstract}

\pacs{\QQ{05.40.Fb (Random walks and Levy flights), 82.70.Dd (Colloids)}}

\maketitle

\section{Introduction}

\footnotetext{\textit{$^{a}$~Condensed Matter Physics Laboratory, Heinrich-Heine-University, Universit\"atsstra{\ss}e 1, D-40225 D\"usseldorf, Germany. E-mail: florian.evers@hhu.de \\ $^{b}$~Institut f\"ur Physikalische Chemie, Universit\"at M\"unster, D-48149 M\"unster, Germany \\ $^{c}$~Fahad Bin Sultan University, SA-71454 Tabuk, Saudi-Arabia}}

The Brownian motion of colloidal particles is one of the classical phenomena in statistical physics \cite{Haw2002,Frey2005,Babic2005,Hanggi2005}. In real situations, the particles often do not move freely, but their dynamics are modified by an external potential \cite{Sokolov2012,Lowen2008,vanBlaaderen2004}. Especially a random potential, and thus Brownian motion in the presence of disorder, leads to interesting transport phenomena \cite{Bouchaud1990, Dean2007}. Up to now, the dynamics in random potentials have been studied mainly by theory and computer simulations \cite{Havlin1987, Isichenko1992, Jack2009, Bernasconi1979, Haus1982, Novikov2011, Scher1973, Schmiedeberg2007, Lacasta2004, Dyre1995}. Theoretical models include the random barrier model \cite{Bernasconi1979}, the random trap model \cite{Haus1982}, the random walk with barriers \cite{Novikov2011} and the continuous time random walk \cite{Scher1973} as well as studies of diffusion in a rough potential \cite{Zwanzig1988} and in materials with defects like zeolites \cite{Chen2000}. In particular, the long-time limit has been investigated for different realizations of random potentials \cite{Bouchaud1990,Dean2007}. In contrast, less is known on the intermediate regime and the time needed to reach the long-time limit. To our knowledge, only very few systematic experimental tests of theoretical and simulation predictions have been performed \cite{Sciortino2005, Hanes2012, Hanes2012a}. Nevertheless, the theoretical predictions have been applied successfully to experimental data and the concept of particles diffusing through an energy landscape has proven to be very useful in understanding very different phenomena. This includes particle diffusion in inhomogeneous media (e.g.~single molecule dynamics in porous gels \cite{Dickson1996} or in cells \cite{Weiss2004,Tolic2004,Banks2005}), the dynamics on rough surfaces \cite{Naumovets2005,Barth2000}, the dynamics of particles moving along the walls between magnetic domains \cite{Tierno2010,Sciortino2005}, the dynamics of independent charge carriers in a conductor with impurities (in the parameter range where conduction can be modeled as a classical process) \cite{Bystrom1950,Heuer2005}. \QQQ{In particular, random potentials with a Gaussian distribution of energy levels have been suggested for different systems \cite{Sciortino2005,Dean2007,Sengupta2005}.} Furthermore, some processes can be represented by a trajectory in the systems' configuration space, for example vitrification leading to glassy systems \cite{Heuer2008, Debenedetti2001, Lubchenko2007, Angell1995, Poon2002, Megen1998, Heuer2005a} or protein folding \cite{Best2011, Dobson1998, Bryngelson1995, Durbin1996, 
Dill1997,Winter2007}. Often diffusion in a \QQQ{random} potential energy landscape represents a crude approximation only, but it can nevertheless provide a useful first description of the effect of disorder on the dynamics \cite{Bouchaud1990,Wolynes1992}. Disorder may modify the value of the diffusion coefficient or it may alter Brownian motion leading to anomalous diffusion. Which effect dominates depends not only on the specific process, but also on the time scale of interest.

An external potential can be imposed on a polarizable colloidal particle by exposing it to a light field \cite{Jenkins2008a, Ashkin1997, Molloy2002,Bowman2013}. Light exerts different forces on particles, if their refractive index differs from (typically exceeds) that of the solvent: a scattering force or `radiation pressure', which pushes particles along the laser beam, and a gradient force, which attracts particles toward regions of high light intensity \cite{Ashkin1997, Molloy2002,Bowman2013}. A classical application of this effect are optical tweezers which are used to trap individual particles by a tightly focused laser beam \cite{Molloy2002, Ashkin1997, Hanes2009, Grier2003, Dholakia2008, Bowman2013}. Furthermore, above a certain light intensity, a periodic light field can induce a disorder-order transition in a two-dimensional charged colloidal system, known as light-induced freezing. If the intensity is increased further, the induced crystal melts into a modulated liquid; this process is called light-induced melting \cite{Loudiyi1992,Wei1998, Bechinger2000}. In addition to the particle arrangement, the particle dynamics can be affected by periodic \cite{Dalle-Ferrier2011} and random \cite{Hanes2012} light fields, resulting in anomalous diffusion. Light fields hence provide a means to manipulate the spatial arrangement and dynamics of colloidal particles. 

Recently, we experimentally realized one-dimensional random energy landscapes \cite{Hanes2009,Hanes2012} and periodic potentials \cite{Jenkins2008a,Dalle-Ferrier2011} using laser light fields  and studied the dynamics of individual particles in these potentials. Here, this is extended to the dynamics of individual colloidal particles in two-dimensional random potentials.
In our experiments and simulations, the values of the two-dimensional random potential were drawn from a Gaussian distribution, whose width $\varepsilon$ represents the degree of roughness of the potential and, in the experiments, was controlled by the laser power $P$. The static properties of the potential were determined quantitatively. Furthermore, the trajectories of individual particles in this potential were followed using video microscopy \cite{Crocker1996,Habdas2002,Kreuter2012} and compared to our simulation results. The dynamics were characterized by, e.g., the time-dependent diffusion coefficient, the mean squared displacement (MSD), the non-Gaussian parameter, and the van Hove function. The dynamics are initially diffusive but then, at intermediate times, show an extended subdiffusive regime before diffusive behaviour is reestablished at very long times. Our findings are compared to the particle dynamics in one-dimensional random potentials \cite{Hanes2012, Hanes2012a} and periodic potentials \cite{Dalle-Ferrier2011}. In two-dimensional potential energy landscapes, particles can bypass large barriers. Therefore, the particle dynamics are controlled by minima and saddle points instead of minima and maxima. Moreover, compared to periodic potentials, the barriers have different heights, which significantly affects the particle dynamics.

\section{Materials and methods}

\subsection{Sample preparation}

Each sample consisted of surfactant-free sulfonated polystyrene particles with a radius $R=1.4\;\micron$ and polydispersity 3.2~\% (Interfacial Dynamics Microspheres \& Nanospheres) suspended in heavy water (D$_2$O), so that the particles cream rather than sediment. Stock solutions of the particles were diluted to result in an area fraction of the creamed sample, $\sigma < 0.10$, which represents a compromise between negligible particle--particle interactions and reasonable statistics. Area fractions were estimated from micrographs according to $\sigma = \pi R^2 N / A$ with $N$ and $A$ being the number of particles and the area covered by the light field, respectively. 

The heavy water (D$_2$O) was de-ionised by stirring with ion exchange resin to increase the particle--glass repulsion and thus reduce the fraction of particles sticking to the glass surface. To further reduce sticking, all glassware was sonicated in 2\% Helmanex II solution at about 60~$^{\circ}$C and then rinsed with Millipore water and dried in air prior to use. Each sample cell was constructed from a microscope slide and three cover slips, two used as spacers (number 0 with thickness $0.085 - 0.13$~mm, supplied by VWR) with a gap between them and the third on top to create a narrow capillary (number 1 with thickness $0.13 - 0.16$~mm, supplied by VWR) \cite{Jenkins2008}. Thin cover glasses were used as spacers to allow imaging of the creamed particles using a high resolution objective with a working distance of 0.13~mm. The sample chamber was filled using capillary action and subsequently sealed with UV glue.

\subsection{Light field generation}
\label{sec:lightField}

The set-up contains a laser with a wavelength of 532~nm (Ventus 532-1500, Laser Quantum). Its beam is expanded and then reflected from a spatial light modulator (Holoeye 2500-LCR). Subsequently, it is directed through two telescopes to reduce its diameter and reflected off three mirrors to steer it through an inverted microscope (Nikon Eclipse 2000-U) into the sample \cite{Hanes2009, Hanes2012, Zunke2013}. One of the mirrors is a dichroic mirror to introduce the beam into the microscope beam path and to use the microscope objective (60$\times$ oil immersion, numerical aperture NA~1.4, Nikon) to image the light field into the sample plane. The beam passes upwards through the sample and hence, due to radiation pressure, pushes the particles against the top of the cell, which reinforces the creaming of the particles. A notch filter in the imaging path prevents laser light from reaching the ocular or camera. To aid alignment, the notch filter can be removed and the sample replaced by a mirror, so that the light intensity distribution in the sample plane can be imaged using the microscope. 

\begin{figure}[t!]
    \centering
   \includegraphics[width=4cm]{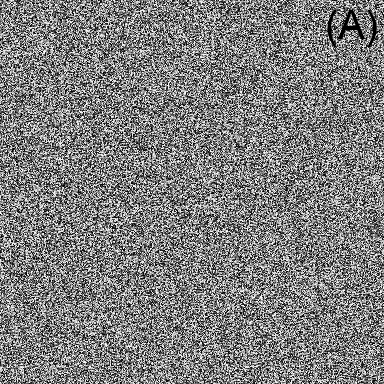} 
   \includegraphics[width=4cm]{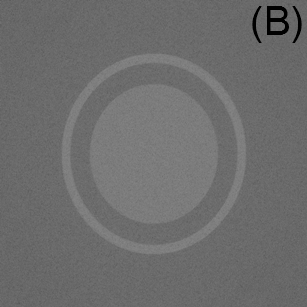} 
   \includegraphics[width=4.2cm]{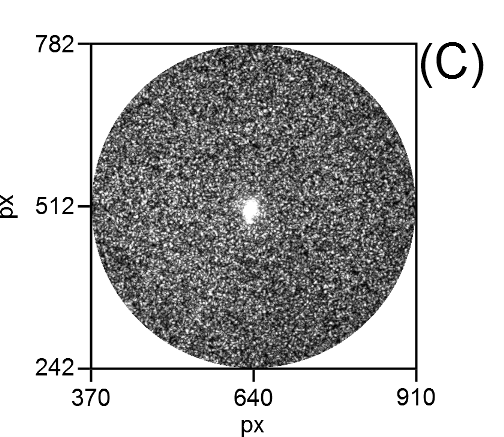}   
   \includegraphics[width=4.2cm]{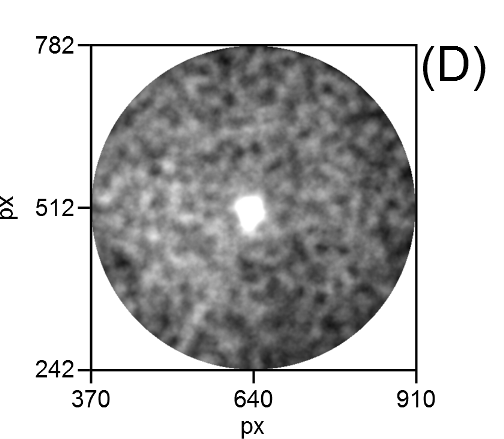}
  \caption{(A) Kinoform calculated by applying the Gerchberg-Saxton algorithm to a homogeneous disc surrounded by a ring and (B) its Fourier transform. (C) Micrograph of the observed intensity $I(x,y)$ of the disc taken at very low laser power $P \leq 0.2$~mW. (D) Potential $U(x,y)$ as experienced by a point-like test particle obtained by convoluting $I(x,y)$ with the volume of a spherical particle with radius $R=1.4\;\micron\;\hat = \; 12.7$~px.}
  \label{fgr:lightfield_holo}
  \end{figure}

A kinoform (phase hologram) was calculated using the Gerchberg-Saxton iterative algorithm \cite{Gerchberg1972} (Fig.~\ref{fgr:lightfield_holo}A) and displayed in the centre of the spatial light modulator. The kinoform corresponds to a homogeneous  disc surrounded by a ring to prevent particle movements into and out of the disc. The Fourier transform of the kinoform is, as expected, a homogeneous disc surrounded by a ring (Fig.~\ref{fgr:lightfield_holo}B). In order to account for the angle at which the laser impinges on the spatial light modulator ($22.5^{\circ}$), the disc and ring are a factor $1/\cos{(22.5^{\circ})}=1.08$ taller than they are wide \cite{Hanes2009,Hanes2012}. The observed light field intensity $I(x,y)$ (Fig.~\ref{fgr:lightfield_holo}C) corresponds to the disc of the Fourier transformed kinoform. Indeed, the illumination is overall flat but, crucially, has some fluctuations due to the finite size and pixelation of the light modulator \cite{Hanes2012}. These fluctuations are exploited in the following. Furthermore, there is a bright 0$^{\mathrm{th}}$-order peak in the centre. Using this peak, a particle was trapped and used to monitor any drift of the set-up \cite{Hanes2012}. Global drifts were found to be negligible during individual measurements (up to 4~h).
 
\subsection{Video microscopy and particle tracking}

The samples were observed using the inverted microscope. Micrographs were recorded using a CMOS camera (PL-B742F, Pixelink). Particle coordinates were extracted from the time series of micrographs and the trajectories determined using IDL routines \cite{Crocker1996}. \QQ{To allow for an unambiguous reconstruction of the trajectories, the distance particles move between two images was required to be much smaller than the average interparticle distance and thus limited to $1.2\,R$.} \Q{Furthermore, care was taken that particles do not approach each other or the boundary closely such that particle--particle and particle--boundary interactions can be neglected.} Typical measurement times were $2$ to $3$~h. Particles which were stuck to the glass were identified by comparing the particles' short-time friction coefficient $\xi_i$, i.e.~the inverse mobility, determined from the mean squared displacement, to the expected bulk value $\xi_0=6 \pi \eta R$ with the solvent viscosity $\eta = 1.19 \times 10^{-3}$~Pa$\,$s at room temperature. Particles with $\xi_i > 20\, \xi_0$ were declared stuck and removed from the analysis. Typically, one particle was stuck to the glass in the field of view, which contained about 20 particles. 

For identical conditions, measurements at different positions in the sample yielded very similar results, despite slightly different particle area fractions $\sigma$. This reproducibility allowed us to average several independent measurements of equal recording time $T_{\text{exp}}$ to improve statistics. 

\subsection{Monte Carlo simulations}

The Monte Carlo simulations were performed on a $4096 \times 4096$ square lattice with the lattice points separated by a distance $\Delta s$ in both directions, where we have set $\Delta s = 1$. The potential values at the lattice points, $\tilde{U}(x,y)$, were produced using a Box-Muller algorithm generating numbers which are Gaussian distributed with zero mean and standard deviation $\tilde{\varepsilon}$. The potential $\tilde{U}(x,y)$ was convoluted with the particle volume to obtain the potential $U(x,y)$ felt by a point-like test particle
\begin{eqnarray}
U(x,y) = \frac{\sum\limits_{k} \sum\limits_l \tilde{U}(x{-}k\Delta s,y{-}l\Delta s) \; a(k,l)}{\sqrt{\sum\limits_k \sum\limits_l a^2(k,l)}}
\end{eqnarray}
where the double sum runs over the projected particle, i.e.~$k^2+l^2 \le m^2$ with $k\Delta s$ and $l \Delta s$ the distances from the particle centre in the two directions and $R=m \Delta s$ the radius of the particle. The volume of the particle is represented by
\begin{eqnarray}
a(k,l) = 2\sqrt{(m^2-k^2-l^2)} \;\; .
\end{eqnarray}
As a compromise between negligible discretization effects and viable computation time, we have chosen $m=32$ and thus $-32 \le k,l \le 32$.

\begin{figure}[t]
\includegraphics[clip=false,width=0.950\columnwidth]{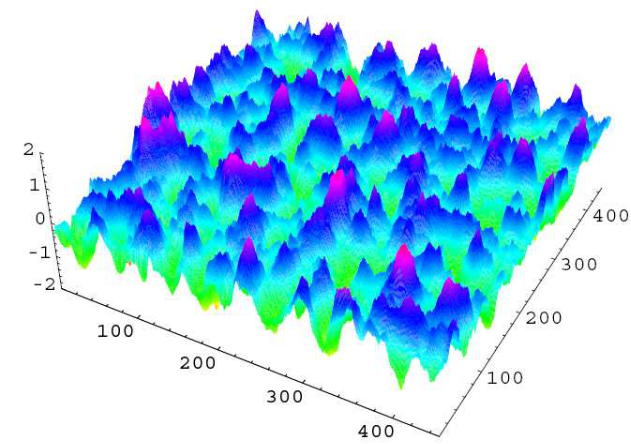}
\caption{Some region of the spatially correlated Gaussian potential energy landscape $U(x,y) / k_{\text{B}}T$ obtained by convolution of a spatially uncorrelated Gaussian energy landscape with the particle volume. It thus reflects the potential felt by a particle (Fig.~\ref{fgr:lightfield_holo}D) and is used in the Monte Carlo simulations.}
\label{fig:sim}
\end{figure}

The convolution leads to a potential $U(x,y)$ (Fig.~\ref{fig:sim}), which is smoother than $\tilde{U}(x,y)$. Its values follow the same Gaussian distribution, albeit with a spatial correlation decaying on the length scale of the particle size. It is supposed to resemble the potential energy landscape experienced by a colloidal particle in the light field (Sec.~\ref{sec:statics}).

Once the potential energy landscape $U(x,y)$ was fixed, a particle was positioned on a randomly chosen lattice point. During the simulation, a direction is chosen randomly and, depending on the energy difference $\Delta U$ to the neighbouring lattice point, the particle is moved in any case if $\Delta U \le 0$, or moved with a finite probability $\exp{(-\Delta U/k_{\text{B}}T)}$ if $\Delta U > 0$ (where $k_{\text{B}}T$ is the thermal energy). By averaging over 1024 different initial positions of the particle, representative averages can be determined. For each Monte Carlo run, the short-time diffusion coefficient $D_0$ and the related Brownian time $t_{\text{B}}=R^2/4D_0$ were calculated. In analogy to the experiment, data were acquired up to $T_{\text{sim}} = 1000\,t_{\text{B}}$. This yielded particle trajectories as in the experiments. Thus, the different parameters, such as the mean squared displacement, were determined as in the experiments, \QQ{including averaging over waiting times} (see below). It turned out that within statistical uncertainty the results for different realizations of the potential energy landscape $U(x,y)$ are identical. As in the experiments, separate simulations were performed for different values of the degree of roughness $0 \, k_{\text{B}}T  \le \varepsilon \le 3 \, k_{\text{B}} T$ to investigate its effect on the dynamics.

\section{Results and discussion}

We studied the behaviour of individual colloidal particles in two-dimensional random potential energy landscapes. At first, the properties of the experimentally created energy landscapes are presented. Then, the particle dynamics in these energy landscapes are discussed and compared to the results of our Monte Carlo simulations and theoretical predictions. Finally, our experimental and simulation results are contrasted with the dynamics in one-dimensional random and periodic potentials.

\subsection{Properties of the optically generated random potential}
\label{sec:statics}

A realization of the light field at very low laser power is displayed in Fig.~\ref{fgr:lightfield_holo}C. The light field interacts with polarizable particles \cite{Ashkin1997, Molloy2002, Bowman2013}. The polarizable particle volume is taken into account by convolving the local light intensity $I(x,y)$ with the particle volume. The effect of the light field on the particle is then represented by an external potential $U(x,y)$ as felt by a point-like test particle (Fig.~\ref{fgr:lightfield_holo}D).

\begin{figure}[t!]
  \centering
   \includegraphics[height=6cm]{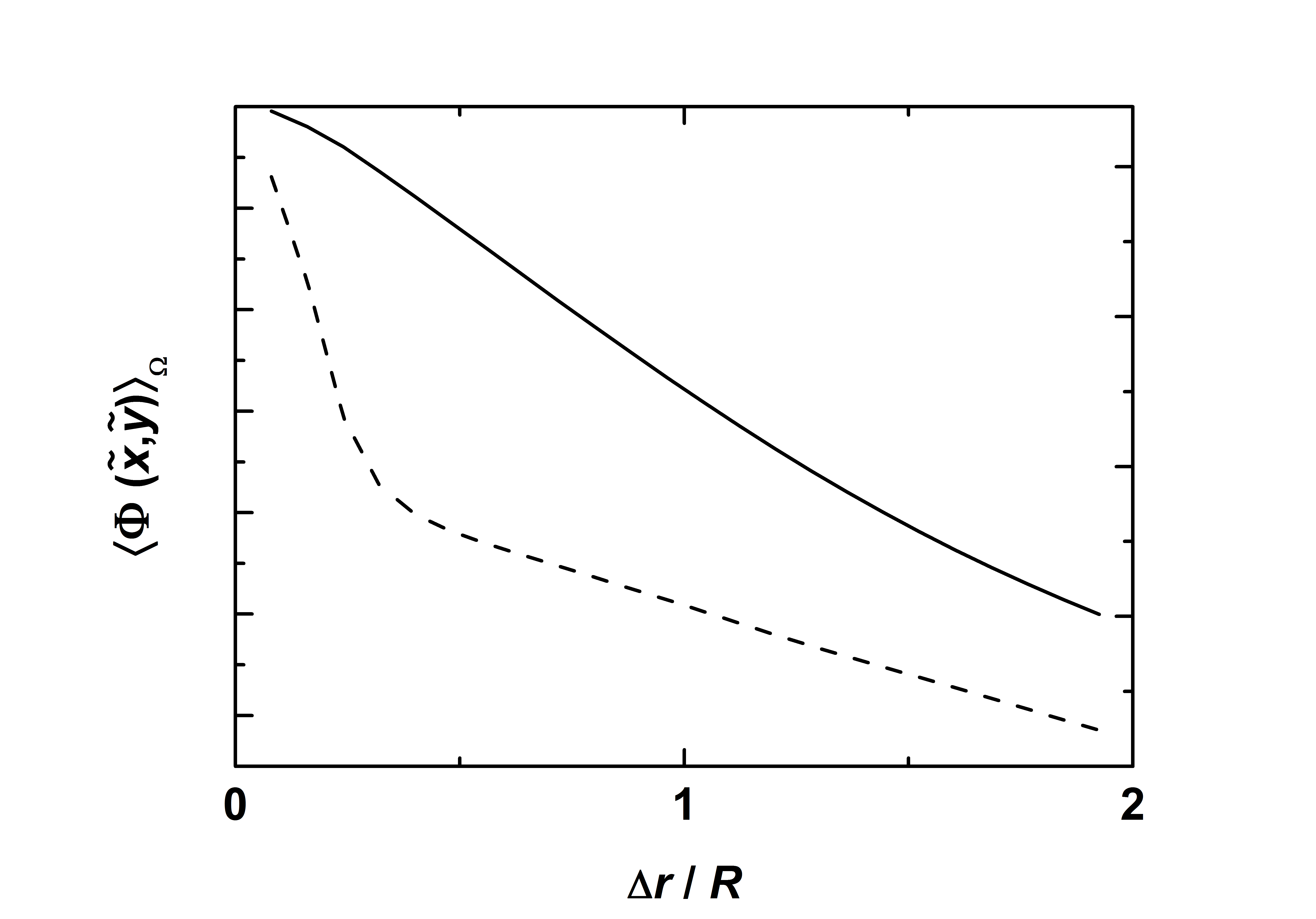}
  \caption{Azimuthally averaged spatial correlation function  $\left\langle \Phi(\tilde{x},\tilde{y}) \right\rangle_{\mathrm{\Omega}}$ of the laser intensity $I(x,y)$ (dashed line) and of the potential energy landscape felt by a point-like test particle $U(x,y)$ (solid line) vs.~the distance $\Delta r$ as determined from Fig.~\ref{fgr:lightfield_holo}C and D, respectively.}
  \label{fgr:corrf}
\end{figure}

To determine the characteristic length scales of the light field intensity $I(x,y)$ and of the potential felt by a point-like test particle $U(x,y)$, the spatial correlation functions were determined and their azimuthal average $\left\langle \Phi(\tilde{x},\tilde{y}) \right\rangle_{\mathrm{\Omega}}$ calculated. The spatial correlation of the light field intensity $I(x,y)$ decays on a short length scale compared to the particle size. However, the convolution with the particle volume introduces a length scale, namely the particle diameter $2R$. The spatial correlation of the potential $U(x,y)$, which was similarly determined, indeed decays on a characteristic length of $2R$ (Fig.~\ref{fgr:corrf}).

\begin{figure}[t!]
  \centering
   \includegraphics[height=6cm]{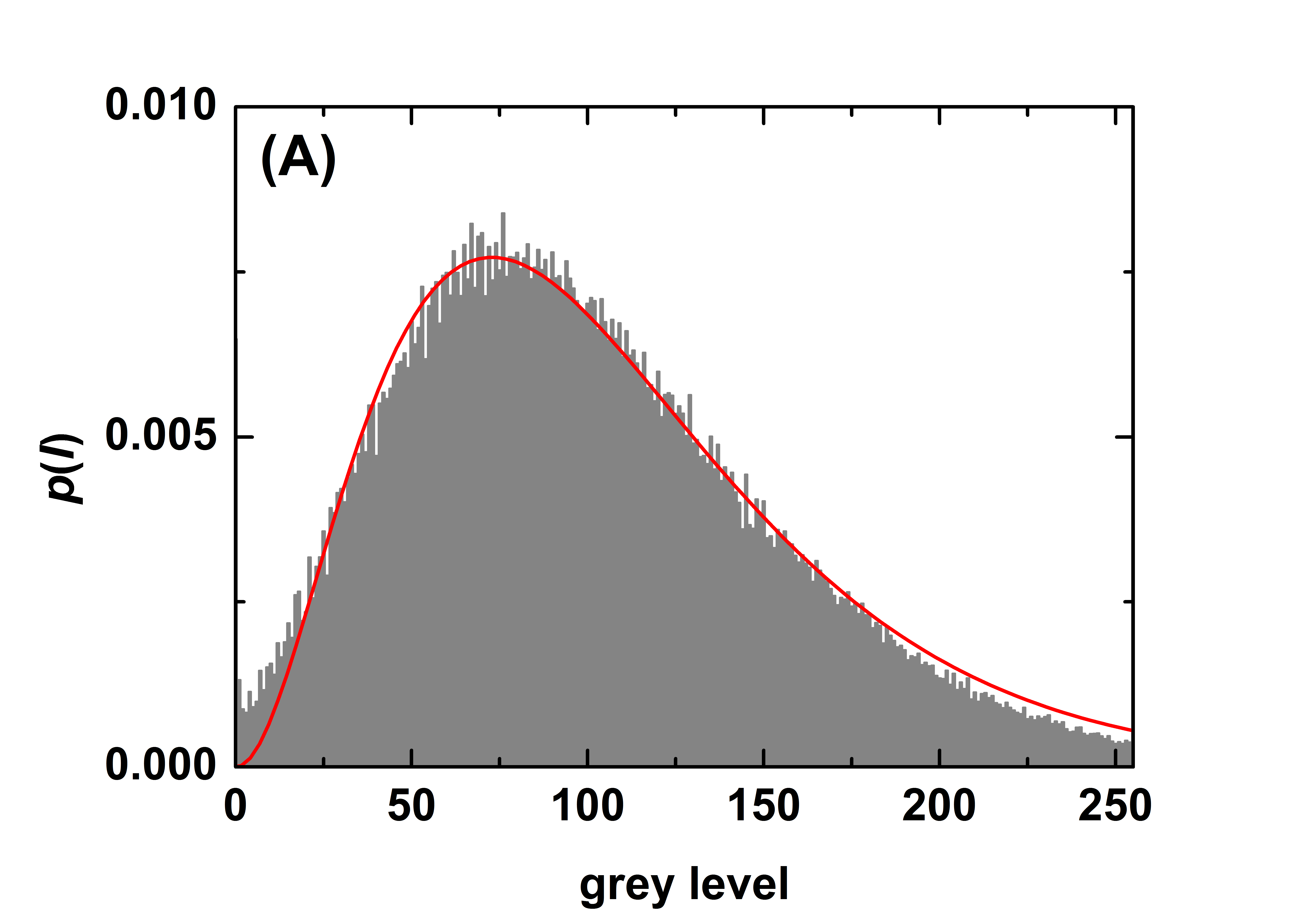}
   \includegraphics[height=6cm]{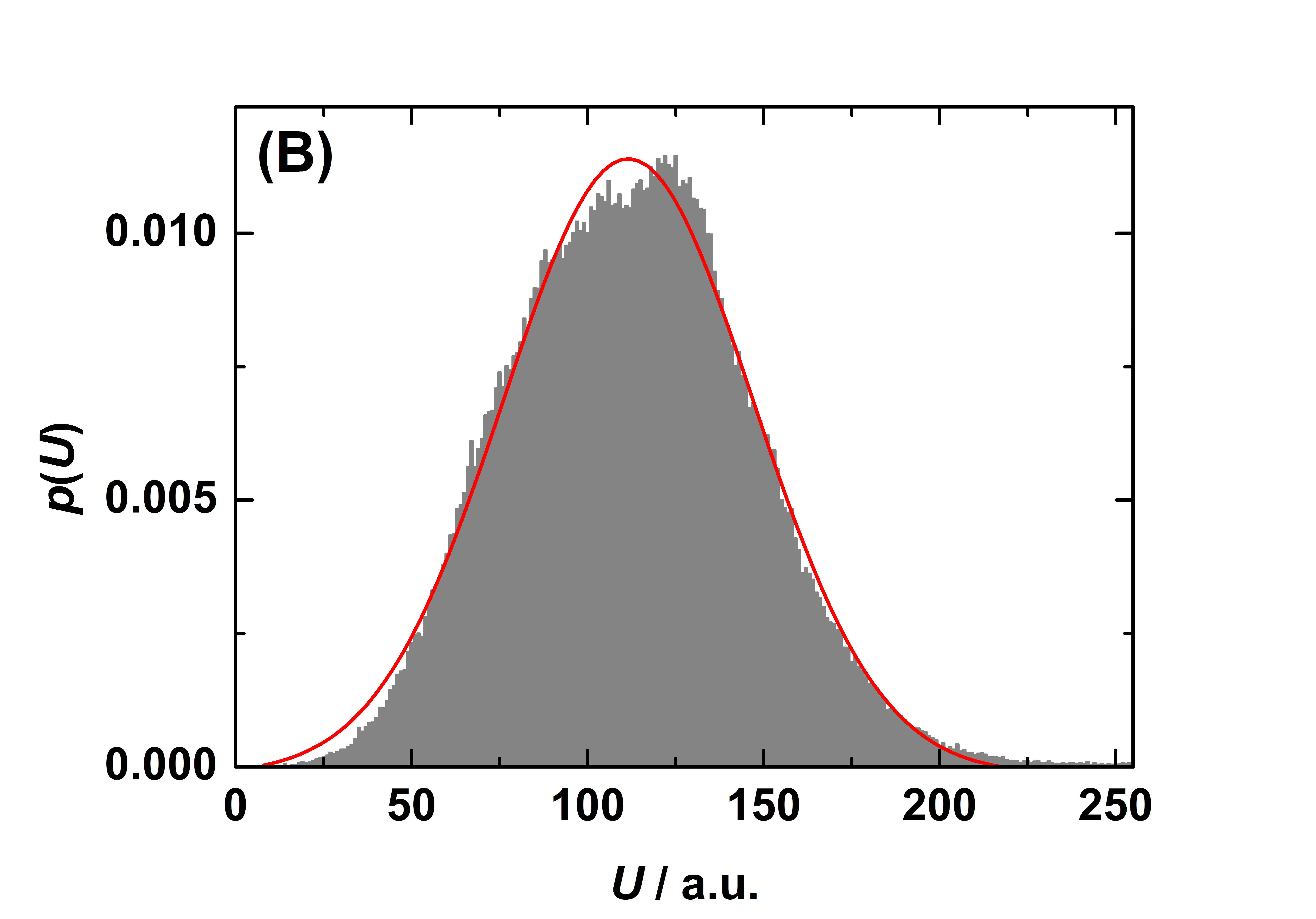}
  \caption{Distribution of (A) values of the intensity of the light field, $p(I)$, and (B) values of the potential as felt by a point-like test particle, $p(U)$, based on the observed intensity $I(x,y)$ and potential $U(x,y)$ shown in Fig.~\ref{fgr:lightfield_holo}C and D, respectively. Red lines are fits based on a Gamma and Gaussian distribution, respectively.}
  \label{fgr:distrib}
\end{figure}

\begin{figure}[t!]
  	\centering
   \includegraphics[height=6cm]{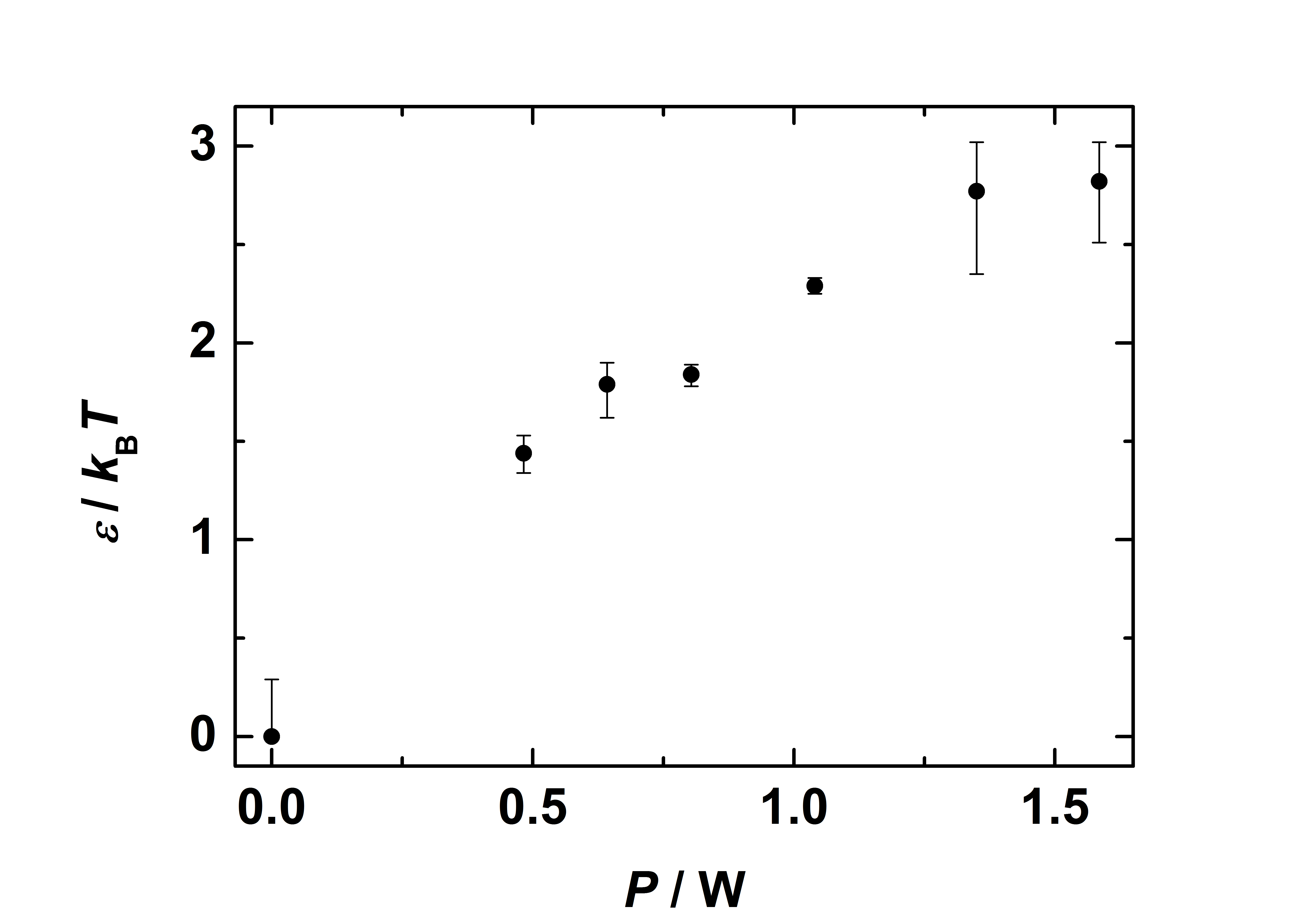}
  \caption{Standard deviation $\varepsilon$ of the distribution of potential energy values, $p(U)$, as a function of laser power $P$.}
  \label{fgr:eps_scale}
\end{figure}

Based on the observed light intensity $I(x,y)$ and potential energy landscape $U(x,y)$ (Fig.~\ref{fgr:lightfield_holo}C,D), the distributions of the light intensity values $p(I)$ and potential values $p(U)$ were determined (Fig.~\ref{fgr:distrib}). The distribution $p(I)$ follows the probability density function of a Gamma distribution \cite{Goodman2010} 
\begin{eqnarray}
		f_{\Gamma}(I) = \frac{b^k}{\Gamma(k)} I^{k-1} e^{-bI},
\end{eqnarray}
where $I \ge 0$, $\Gamma(k)$ is the Gamma function \QQ{and $b$ the scale parameter}. A fit to the experimental $p(I)$ yielded \QQ{a shape parameter} $k = 3.1 \pm 0.1$ (Fig.~\ref{fgr:distrib}A), corresponding to a 3D speckle pattern \cite{Goodman2010, Douglass2012}. The distribution $p(U)$ can be described by a Gaussian distribution
\begin{eqnarray}
	f_{\text{G}}(U) = \frac{1}{\sqrt{2 \pi \varepsilon^2}} e^{-\frac{(U-\langle U \rangle)^2}{2 \varepsilon^2}}
\end{eqnarray}
with the average $\langle U \rangle$ and width or standard deviation $\varepsilon$ (Fig.~\ref{fgr:distrib}B). Due to the convolution with the particle volume, $U(x,y)$ represents a weighted average of several independent (random) values of $I(x,y)$ and thus $p(U)$ has a significantly reduced width compared to $p(I)$. The width $\varepsilon$ characterizes the degree of roughness of the random potential $U(x,y)$, which is controlled by the laser power $P$, but cannot easily be determined experimentally. Thus, to establish a quantitative relation between the roughness $\varepsilon$, used in the simulations, and the laser power $P$, applied in experiments, the experimental potential energy landscape was calibrated. This was achieved by a direct comparison of the experimental and simulation results, namely of the time-dependent diffusion coefficient $D(t)$ at very short and long times (Sec.~\ref{sec:dynSim}). The calibration resulted in an approximately linear relation between $\varepsilon$ and $P$, which might saturate for large $P$ (Fig.~\ref{fgr:eps_scale}).

\subsection{Dynamics in the random potential -- experiments}
\label{sec:dyn}

\begin{figure}[t!]
	\centering
   \includegraphics[height=6cm]{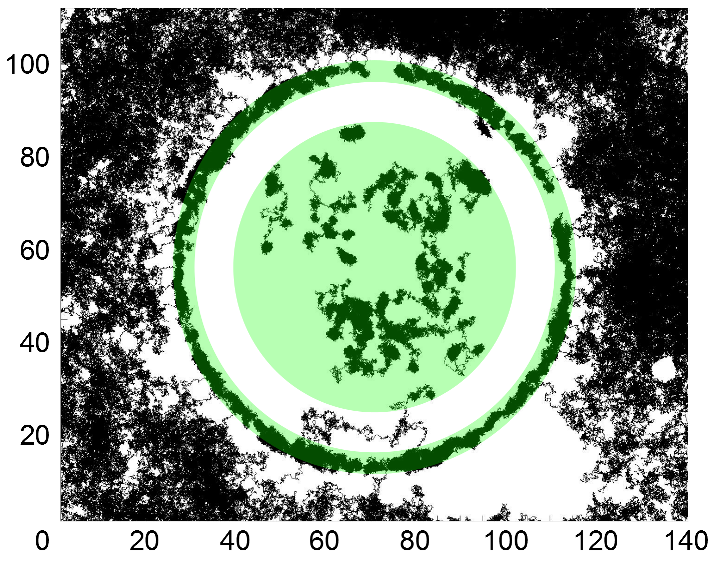}   
  \caption{Trajectories of particles undergoing diffusion in a two-dimensional plane, part of which contains a random potential (green background) which is separated by a barrier (white/green rings) from the surroundings (white background). Particle radius $R=1.4~\micron$, particle surface fraction $\sigma=0.04$, laser power $P=1.32$~W corresponding to a standard deviation $\mathrm{\varepsilon}=2.8~k_{\text{B}}T$, and a recording time $T_{\text{exp}}=3.8$~h. Coordinates are given in $\micron$.}
  \label{fgr:traj}
\end{figure}

The effect of two-dimensional random energy landscapes on the particle dynamics is qualitatively illustrated in Fig.~\ref{fgr:traj}. Outside the light field (white background), particles undergo free diffusion, exploring a large area. This region is separated by a large barrier (white/green rings) from the two-dimensional random light field (green disc). Within the random potential, the excursions of the particles are limited and hence the particle dynamics are slowed down. The particles remain longer at some positions, which correspond to local minima of the potential. For a potential with a larger degree of roughness $\varepsilon$, i.e.~a larger width of $p(U)$, this effect is more pronounced with particles being more efficiently trapped and hence exploring a smaller region.

Based on the particle trajectories, different statistical properties were computed to characterize the particle dynamics. We found identical behaviour along the $x$- and $y$-directions as expected for an isotropic system. The dynamical properties were hence determined as a function of the distance, $\mathrm{\Delta} r = [(\mathrm{\Delta} x)^2+(\mathrm{\Delta} y)^2]^{1/2}$, where distances are scaled by the particle radius $R=1.4\;\micron$ and times by the Brownian time $t_{\text{B}}=R^2/(4D_0)=(6.4 \pm 0.1)$~s with $D_0$ experimentally determined in the absence of a random potential, i.e.~$\varepsilon=0$, but in the vicinity of the water--glass interface. This renders the data independent of the specific experimental conditions, \QQ{except for a radiation pressure effect (Sec.~\ref{sec:dynSim}).} Moreover, the statistical properties were obtained by averaging over different particles, which are well separated and thus non-interacting, and over waiting times $t_0$ to improve statistics. Since, initially, the occupancy of energy levels was homogeneous but tended toward a Boltzmann distribution in the course of the experiment, the average over waiting times depends on the total measurement time $T_{\text{exp}}$, which was $T_{\text{exp}} \approx 1000 \, t_{\text{B}}$.

\begin{figure}[t!]
  	\centering
   \includegraphics[width=8cm]{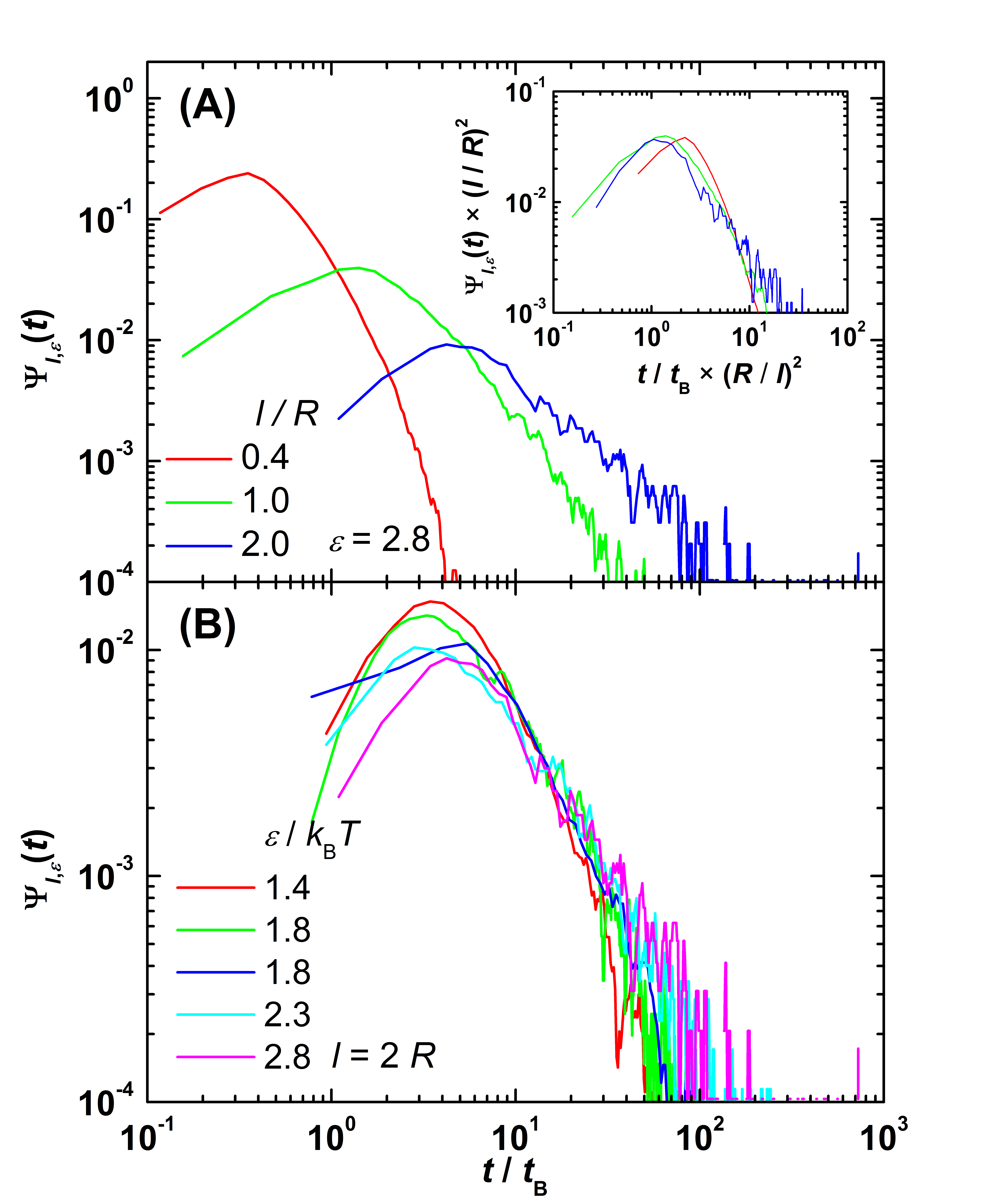}
  \caption{Particle residence time distribution $\Psi_{l,\varepsilon}(t)$ representing the probability that it takes a particle a time $t$ to travel at least a distance $l$ in a random potential with standard deviation $\varepsilon$. All curves are smoothed by a moving five-points average. (A) $\Psi_{l,\varepsilon}(t)$ for different length $l/R$ (as indicated) and $\varepsilon=2.8~k_{\text{B}}T$, scaled plot as inset. (B) $\Psi_{l,\varepsilon}(t)$ for $l/R=2$ and different $\varepsilon$ (as indicated).} 
  \label{fgr:waitingtime}
\end{figure}

Depending on the particle positions, the particles experience various potential values $U(x,y)$ and are trapped for different times, reflecting the different heights of the saddle points to the neighbouring minima. The time $t$ required to explore at least a distance $l$ in a potential with roughness $\varepsilon$ has been determined and the particle residence time distribution $\Psi_{l,\varepsilon}(t)$ calculated. To explore a distance $l$ by free diffusion with diffusion coefficient $D_0$, on average the time $t=l^2/(4D_0)$ is required. To explore larger distances $l$ and/or in the presence of a random potential, on average larger times are required. For short distances $l < 2R$, i.e.~within a minimum, $\Psi_{l,\varepsilon}(t)$ does not significantly depend on $\varepsilon$ \QQ{but depends on the distance $l$ (Fig.~\ref{fgr:waitingtime}A). The $l$ dependence is mainly governed by the longer time required to diffuse a larger distance $l$ as shown by a rescaling assuming diffusive motion (Fig.~\ref{fgr:waitingtime}A, inset). In contrast,} to travel a distance of at least $2R$, which corresponds to the typical minimum-minimum separation (Fig.~\ref{fgr:corrf}), \QQ{in general requires to cross a barrier or saddle point, whose average height depends on $\varepsilon$. Accordingly, $\Psi_{l,\varepsilon}(t)$ depends on the roughness $\varepsilon$ (Fig.~\ref{fgr:waitingtime}B) and the mean residence time exceeds the average time $t = 4 t_{\text{B}}$ required to diffuse $2R$ in the absence of a potential.}

\begin{figure}[t!]
  	\centering
   \includegraphics[width=7cm]{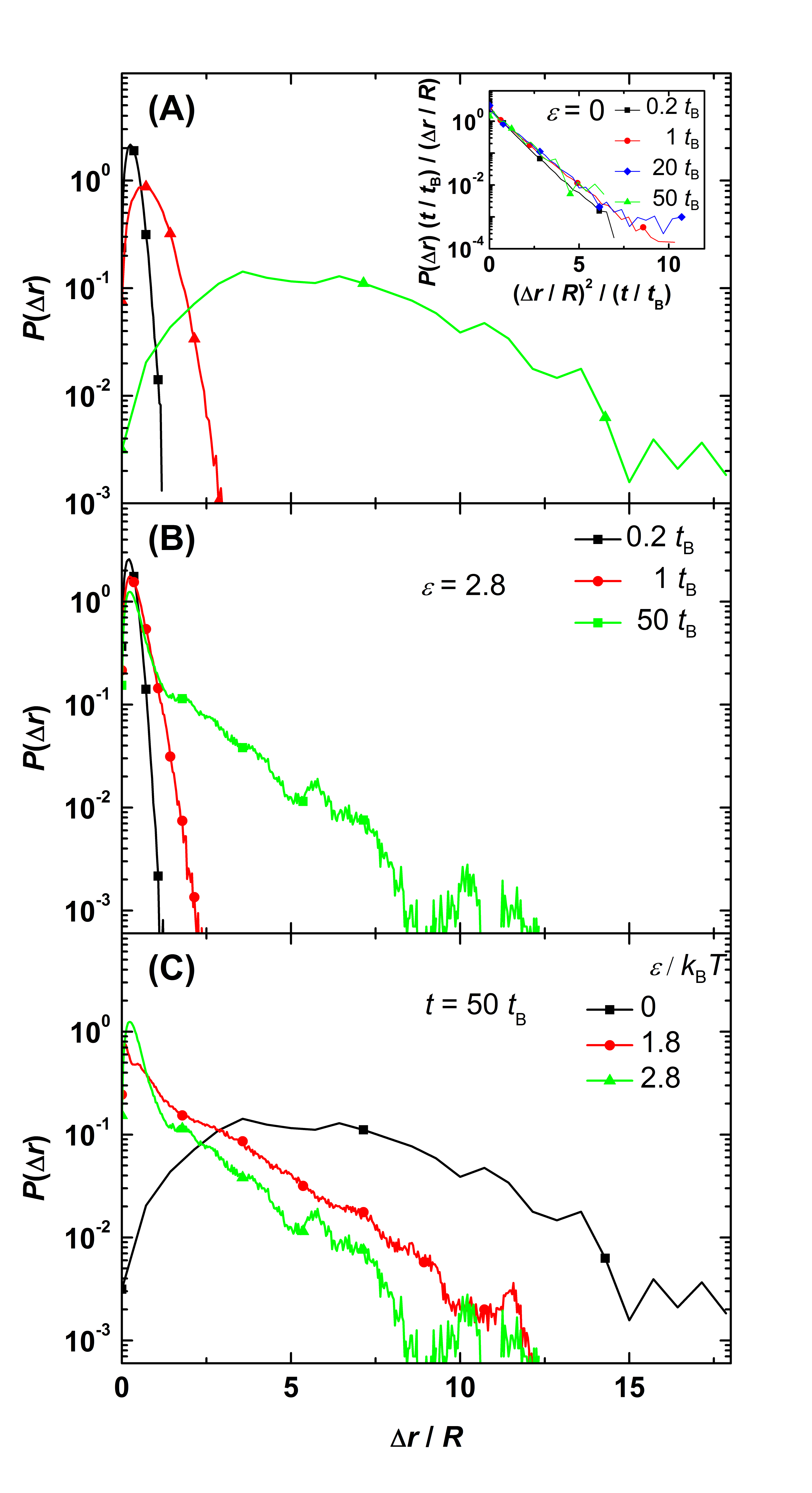}
  \caption{Distribution of particle displacements $\Delta r$ within time $t$, $P(\Delta r, t)$ (A) in the absence of a potential ($\varepsilon=0$), i.e.~for free diffusion, with the scaled $P(\Delta r,t)$ as an inset, (B) in the presence of a random potential with roughness $\varepsilon = 2.8~k_{\text{B}}T$ for different times $t$ (as indicated) and (C) with different roughnesses $\varepsilon$ (as indicated) for time $t=50\,t_{\text{B}}$.}
  \label{fgr:PDF0}
\end{figure}

The probability distribution of particle displacements $\Delta r$, i.e.~the self part of the van Hove function, $P(\Delta r,t)$, at different delay times $t$ is calculated based on the trajectories by averaging over all waiting times $t_0$ and particles $i$:
\begin{eqnarray}
  P (\Delta r,t) = \left\langle \mathrm{\delta}\left( \Delta r - \left[ r_i(t_0{+}t) - r_i(t_0) \right] \right) \right\rangle_{t_0,i},
\end{eqnarray}
where $r_i(t)$ is the position of particle $i$ at time $t$. In the case of free two-dimensional diffusion, i.e.~without any external potential, $P(\Delta r,t)$ follows a Rayleigh distribution, $P (\Delta r,t) \sim \Delta r/(2D_0 t)\,\exp{\left( - \Delta r^2 / 4D_0 t \right)}$, whose width increases linearly with time $t$ (Fig.~\ref{fgr:PDF0}A). In the presence of a random potential, $P(\Delta r,t)$ changes qualitatively (Fig.~\ref{fgr:PDF0}B). The potential tends to trap the particle so that it explores less space and the distributions $P(\Delta r,t)$ get much narrower. This is more pronounced for longer times, when the dynamics include barrier crossing. Accordingly, at long delay times, the roughness of the potential significantly effects $P(\Delta r,t)$, which becomes narrower with increasing $\varepsilon$ (Fig.~\ref{fgr:PDF0}C).

\begin{figure}[t!]
  	\centering
   \includegraphics[width=8cm]{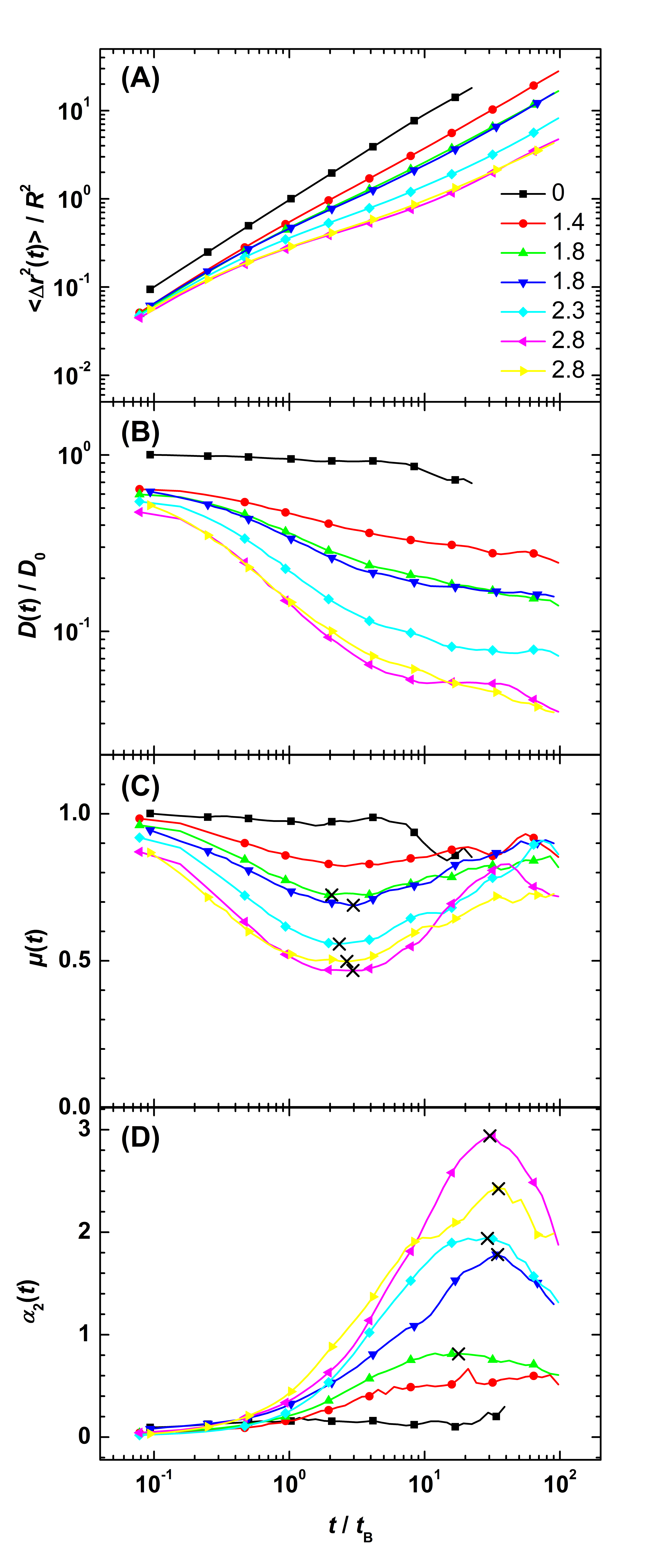}
   	\caption{(A) Normalized mean squared displacement $\left\langle\mathrm{\Delta}r(t)^2\right\rangle/R^2$, (B) normalized diffusion coefficient $D(t)/D_0$, (C) exponent $\mu(t)$ in the relation $\left\langle \mathrm{\Delta} r^2 (t) \right\rangle  \sim t^{\mathrm{\mu}(t)}$ and (D) non-Gaussian parameter $\mathrm{\alpha}_2(t)$ as a function of delay time $t$ normalized by the Brownian time $t_{\text{B}}$ in the presence of a two-dimensional random potential with roughness $\varepsilon$ (as indicated). For clarity, only every fifth data point is plotted as a symbol. Black crosses indicate minima and maxima of $\mu(t)$ and $\mathrm{\alpha}_2(t)$, respectively.}
  \label{fgr:DD0}
\end{figure}

The width of the distribution of particle displacements, $P(\Delta r,t)$, can be characterized by the mean squared displacement (MSD) 
\begin{equation}
  \left\langle \mathrm{\Delta} r^2 (t) \right\rangle = \left\langle \mathrm{\Delta} x^2 (t) \right\rangle + \left\langle \mathrm{\Delta} y^2 (t) \right\rangle \; ,
  \label{eq:msd}
\end{equation}
which is calculated from the particle trajectories according to
\begin{eqnarray}
\langle \mathrm{\Delta} x^2(t) \rangle & = &  \left\langle \left[ x_i(t_0 + t) - x_i(t_0) \right]^2 \right\rangle_{t_0,i} \nonumber \\
 & - & \left\langle \left[ x_i(t_0 + t) - x_i(t_0) \right] \right\rangle_{t_0,i}^2
 \label{eq:msd2}
\end{eqnarray}
and $\langle \mathrm{\Delta} y^2 (t) \rangle$ correspondingly, with the second term correcting for possible drifts. \QQ{In the absence of a potential ($\varepsilon=0$), $\langle \Delta r^2(t) \rangle$ increases linearly with time, as expected for free diffusion (Fig.~\ref{fgr:DD0}A). In the presence of a random potential,} the particle dynamics exhibit three distinct regimes. Both, at short times ($t / t_{\text{B}} \lesssim 0.1$) and long times ($t / t_{\text{B}} \gtrsim 30$), the particle dynamics are diffusive. At small $t$, the diffusive behaviour reflects small excursions within local minima and is thus essentially independent of the roughness $\varepsilon$. Nevertheless, diffusion is reduced compared to free diffusion ($\varepsilon=0$) because laser pressure pushes the particles closer to the water-glass interface and thus reduces their mobility \cite{Pagac1996, Leach2009, Sharma2010}, with only a weak dependence on laser power $P>0$ and hence $\varepsilon>0$. \QQ{Furthermore, the averaging over waiting times $t_0$ (Eqs.~\ref{eq:msd},\ref{eq:msd2}) leads to a reduction of the MSD, especially at short times. This is due to the evolution of the system towards an equilibrium (Boltzmann) distribution which leads to an increasing occupation of deep minima. (Both effects are discussed in more detail in Sec.~\ref{sec:dynSim}.)} For large enough $t$, hopping between minima becomes important and constitutes a random walk. Thus, diffusive behaviour is reestablished at long times, although with a strongly reduced diffusion coefficient. At intermediate $t$, the MSDs exhibit an inflection point, which becomes increasingly pronounced as $\varepsilon$ increases. This subdiffusive behaviour is caused by the particle being trapped in local minima for prolonged times before it escapes to a neighbouring minima. Since there is a wide range of residence times (Fig.~\ref{fgr:waitingtime}), reflecting barriers of different heights, the subdiffusive regime extends over a broad range of times.

From the two-dimensional MSD $\langle \Delta r^2(t) \rangle$, the time-dependent diffusion coefficient $D(t)$ can be calculated according to
\begin{eqnarray}
  D (t) = \frac{1}{2d} \frac{\partial}{\partial t} \left\langle \mathrm{\Delta} r^2 (t) \right\rangle  \;\; ,
  \label{eq:D}
\end{eqnarray} 
where in the present case the dimension $d=2$.
The three regimes discussed above are also reflected in the normalized time-dependent diffusion coefficient $D(t)/D_0$ (Fig.~\ref{fgr:DD0}B). Toward very short times, $D(t)/D_0$ tends toward one (actually slightly below one due to the radiation pressure and the averaging mentioned above and discussed in Sec.~\ref{sec:dynSim}). It strongly decreases at intermediate times to reach a much smaller value $D_\infty$ at long times, where hopping between minima dominates and diffusion is reestablished, reflected in the plateau of $D(t)$ at long times. The asymptotic diffusion coefficient $D_\infty$ was determined experimentally and will be discussed together with the simulation results in Sec.~\ref{sec:dynSim}.

In order to characterize deviations from diffusive behaviour, in particular the subdiffusion at intermediate times, the exponent $\mathrm{\mu}$ in the relation $\left\langle \mathrm{\Delta} r^2 (t) \right\rangle  \sim t^{\mathrm{\mu}(t)}$ is determined from the slope of the MSD in double-logarithmic representation:
\begin{eqnarray}
  \mathrm{\mu}(t) = \frac{\partial \log{\left(\left \langle \Delta r^2(t) \right\rangle \right)}}{ \partial \log{(t)} } \;\, .
\end{eqnarray} 
For free diffusion $\mathrm{\mu}=1$, while $\mathrm{\mu} < 1$ in the case of subdiffusion. The subdiffusive dynamics at intermediate times results in a minimum in $\mu(t)$. It becomes more pronounced with increasing $\varepsilon$, but remains at about the same time (Fig.~\ref{fgr:DD0}C, crosses). In contrast, the diffusive behaviour at short and long times is reflected in the trend of $\mu(t)$ toward one in these two limits.

While the exponent $\mu(t)$ characterizes deviations from diffusive behaviour, the non-Gaussian parameter $\mathrm{\alpha}_2(t)$ quantifies, in the case of one dimension, the deviation of the distribution of particle displacements from a Gaussian distribution. It corresponds to the first non-Gaussian correction \cite{Megen1998}. In two dimensions, it quantifies deviations from a Rayleigh distribution (Fig.~\ref{fgr:PDF0}). Following a previous definition \cite{Vorselaars2007}: 
\begin{eqnarray}
  \mathrm{\alpha}_2(t) = \frac{ \left\langle \mathrm{\Delta} r^4 (t) \right\rangle}{(1+2/d)  \left\langle \mathrm{\Delta} r^2 (t) \right\rangle^2} -1,
\end{eqnarray}
where $\left\langle \mathrm{\Delta} r^4 (t) \right\rangle = \left\langle \mathrm{\Delta} x^4(t) \right\rangle + \left\langle \mathrm{\Delta} y^4(t) \right\rangle + 2\left\langle \mathrm{\Delta} x^2(t) \right\rangle \left\langle \mathrm{\Delta} y^2(t) \right\rangle$ and $\left\langle \mathrm{\Delta} x^4(t) \right\rangle$ and $\left\langle \mathrm{\Delta} y^4(t) \right\rangle$ are defined in analogy to $\left\langle \mathrm{\Delta} x^2(t) \right\rangle$. 
The time-dependence of $\mathrm{\alpha}_2(t)$ also shows three different dynamic regimes (Fig.~\ref{fgr:DD0}D).
At very short and very long times, when the particle dynamics are diffusive, $\mathrm{\alpha}_2(t) \approx 0$, while at intermediate times $\mathrm{\alpha}_2(t)$ develops a peak which becomes more pronounced and moves to larger times with increasing $\varepsilon$. This reflects the broader distribution of barrier heights and hence residence times $\Psi_{l,\varepsilon}(t)$ at larger $\varepsilon$ (Fig.~\ref{fgr:waitingtime}).

\begin{figure}[t!]
	\centering
   \includegraphics[height=6cm]{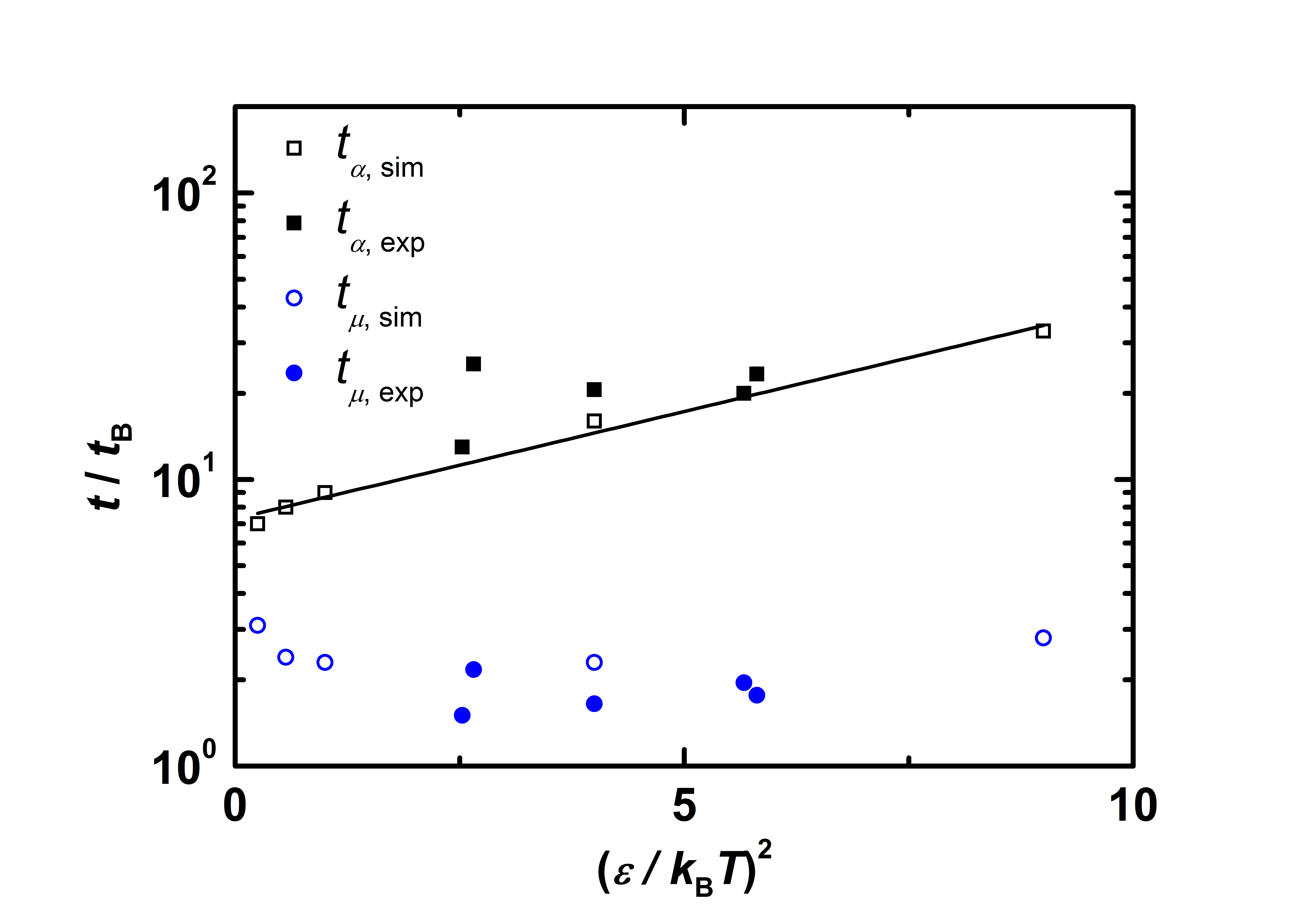}
  \caption{Characteristic times, namely of the minimum in the exponent $\mu(t)$, i.e.~$t_\mu$, and the maximum in the non-Gaussian parameter $\alpha_2(t)$, i.e.~$t_\alpha$, as a function of the degree of roughness $\varepsilon$ of the potential from experiments (filled symbols, corresponding to the crosses in Fig.~\ref{fgr:DD0}C,D \QQ{but taking the radiation pressure effect, as quantified in the inset of Fig.~\ref{fgr:exp_sim}, into account}) and simulations (open symbols). The solid line is a guide to the eye.}
  \label{fgr:tmax}
\end{figure}

All parameters indicate an intermediate time regime characterized by subdiffusive dynamics. In particular, a minimum in the exponent $\mu(t)$ at $t_\mu$ and, at a later time $t_\alpha$, a maximum in the non-Gaussian parameter $\alpha_2(t)$ are observed (Fig.~\ref{fgr:DD0}C,D). While the time $t_\mu$ hardly depends on $\varepsilon$, the maximum in $\alpha_2(t)$ shifts to significantly larger times $t_\alpha$ with increasing $\varepsilon$ (Fig.~\ref{fgr:tmax}). \QQ{The minimum of $\mu(t)$ is reached when diffusion is most efficiently suppressed. This occurs just before a significant fraction of the particles start to escape the minima. This implies relatively shallow minima, which have a similar depth for essentially all $\varepsilon$. Thus the dependence of $t_\mu$ on $\varepsilon$ is very small. On the other hand, the maximum of $\alpha_2(t)$ occurs when the dynamics is maximally heterogeneous, i.e.~some minima have long been left, others only recently and some not yet. This spread increases with $\varepsilon$ and hence does the maximum of $\alpha_2(t)$. Accordingly, to reach this maximally heterogeneous state takes longer and thus $t_\alpha$ increases with $\varepsilon$.}

\subsection{Dynamics in the random potential -- simulations}
\label{sec:dynSim}

\begin{figure}[t!]
  	\centering
   \includegraphics[height=6cm]{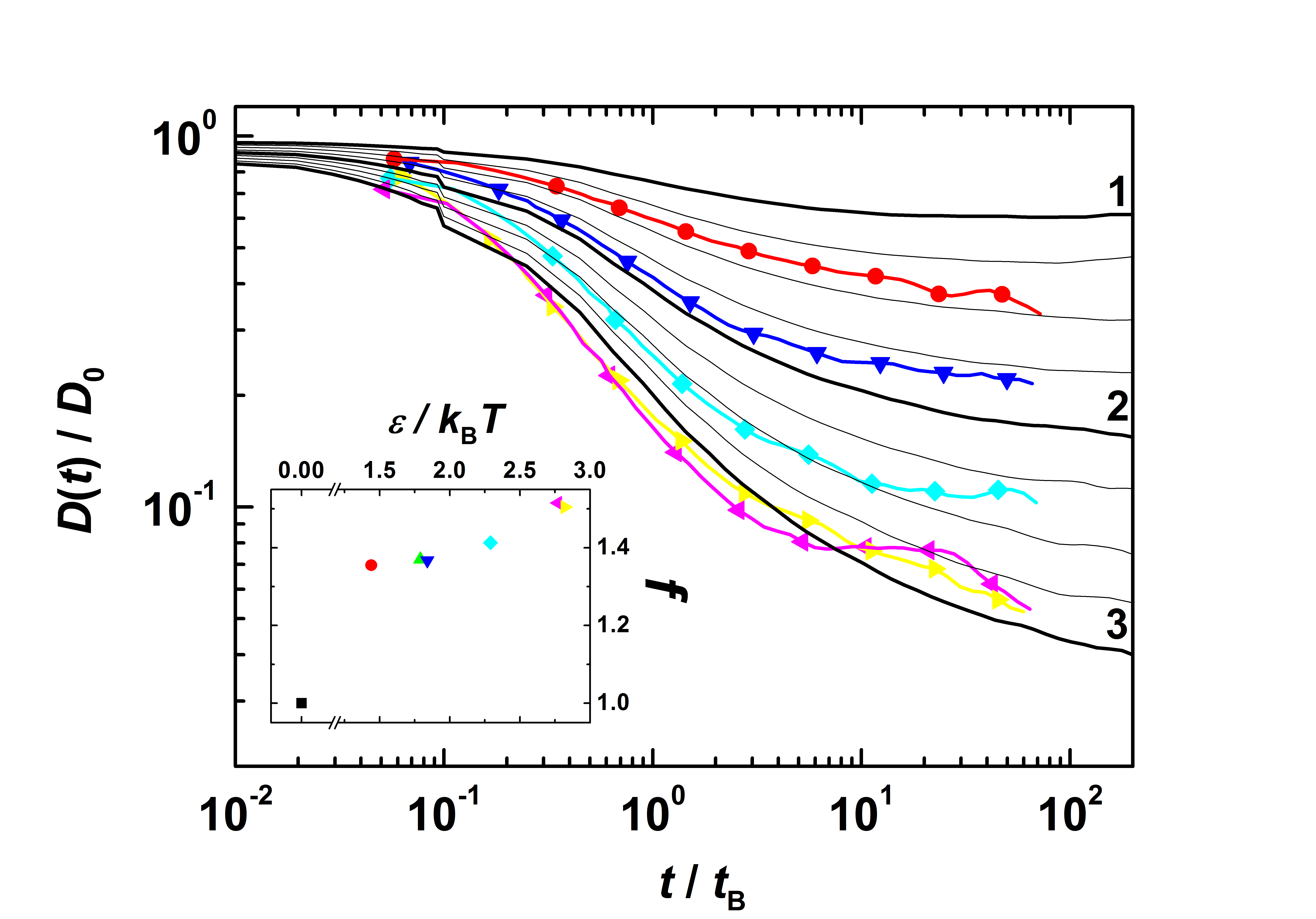}
  \caption{Normalized diffusion coefficient $D(t)/D_0$ as a function of delay time $t/t_{\text{B}}$ for different roughnesses $\varepsilon$ from simulations (black solid lines, \QQQ{for $\varepsilon / k_{\text{B}}T = 1, 1.25, 1.5, 1.75, 2, 2.25, 2.5, 2.75, 3$}) and experiments (coloured lines with symbols as in Fig.~\ref{fgr:DD0}). Here, the experimental data are scaled with an effective diffusion coefficient and effective Brownian time to account for the effect of radiation pressure (see text for details). \Q{The dependence of the scaling factor $f$ on radiation pressure, which is proportional to the laser power $P$ and hence the degree of roughness $\varepsilon$ (Fig.~\ref{fgr:eps_scale}), is shown as inset.}}
  \label{fgr:exp_sim}
\end{figure}

\QQ{The simulations also show three regimes: initially diffusion followed by subdiffusive behaviour and finally again diffusion with a considerably reduced diffusion coefficient $D_\infty$ (Fig.~\ref{fgr:exp_sim}), consistent with our experimental findings (Fig.~\ref{fgr:DD0}).}

\QQ{Already at short times, the diffusion coefficient $D(t)$ is noticeably reduced. The reduction, caused by the random potential, is considerably enhanced by the averaging over waiting times $t_0$ (Eqs.~\ref{eq:msd},\ref{eq:msd2}). As time progresses, the initially homogeneous particle distribution develops into the equilibrium distribution with the energy levels occupied according to the Boltzmann distribution. This implies a growing occupation of deep minima, in which the particles reside for a long time, and hence slower dynamics. With increasing simulation time $T_{\text{sim}}$ (or measurement time $T_{\text{exp}}$), and hence an increasing range of waiting times $0 \le t_0 \le T_{\text{sim}}{-}t$ included in the average, the weight of near-equilibrium distributions with a large fraction of less-mobile particles increases. Hence, the averaging over $t_0$ leads to a smaller mean diffusion coefficient $D(t)$ with the decrease becoming more pronounced as $T_{\text{sim}}$ increases and $t$ decreases. The decrease of $D(t)$ is thus particularly noticeable at short times $t$. Furthermore, the simulation time has to be matched to the measurement time, $T_{\text{sim}} \approx T_{\text{exp}}$, to allow for a meaningful comparison.}

\begin{figure}[t!]
	\centering
	 \includegraphics[height=6cm]{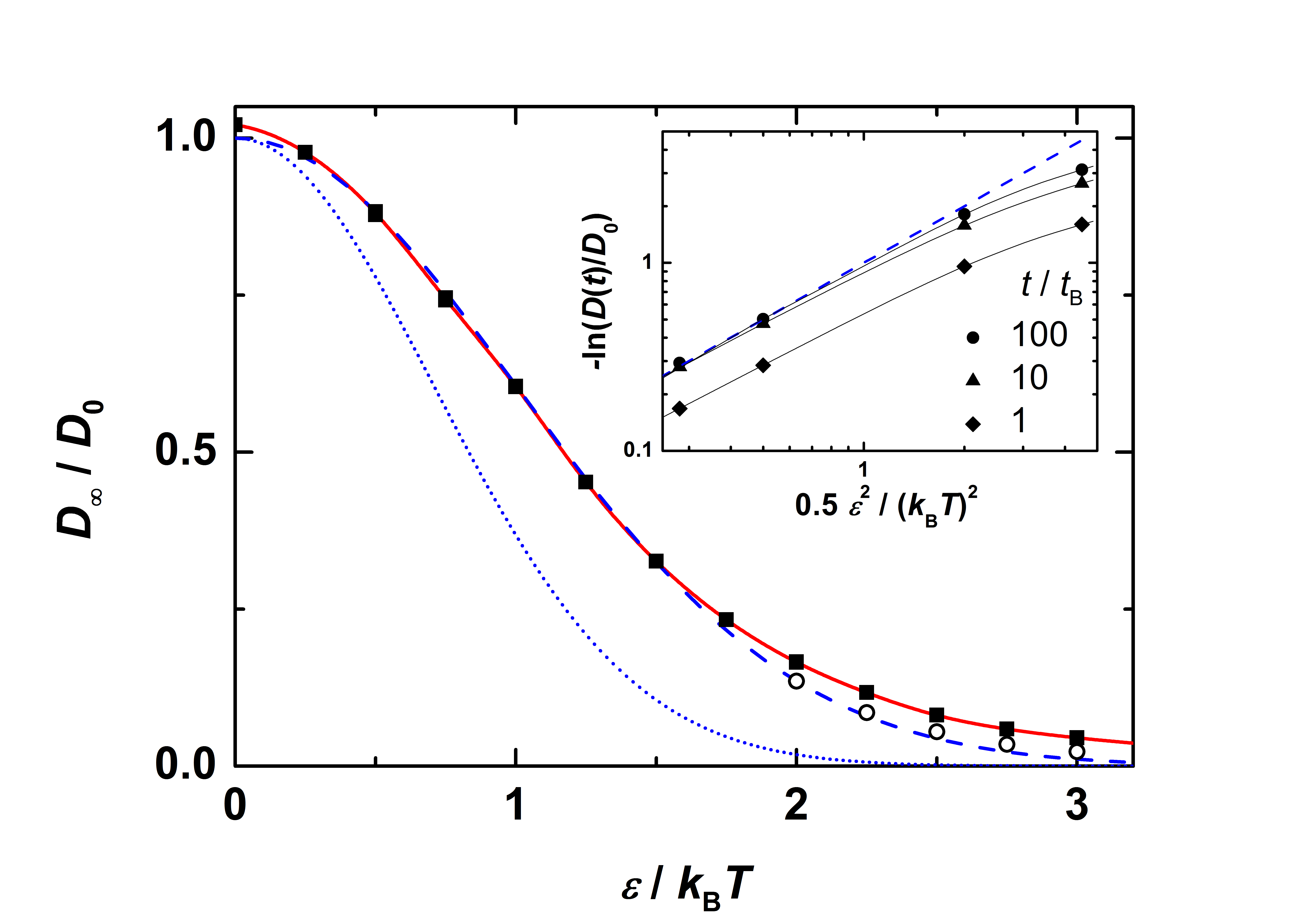}
  \caption{\QQ{Ratio of the long-time diffusion coefficient $D_\infty \approx D(80t_{\text{B}})$ and the diffusion coefficient $D_0$ in the absence of a potential as a function of the degree of roughness $\varepsilon$ as obtained from simulations. \Q{Solid symbols: total simulation time similar to the experimental recoding time, $T_{\text{sim}} \approx T_{\text{exp}}$, open symbols: one order of magnitude longer simulation time, $T_{\text{sim}} \approx 10 \, T_{\text{exp}}$, $D_\infty \approx D(1000 \, t_{\text{B}})$}. The red solid line represents a spline interpolation of the simulation data and the blue dashed and dotted lines the theoretical predictions for a two-dimensional \cite{Dean2007} and one-dimensional \cite{Zwanzig1988} random potential, respectively. The inset shows the ratio $D(t)/D_0$ at different times $t=t_{\text{B}}$, $10\,t_{\text{B}}$ and $100\,t_{\text{B}}$. The black lines are guides to the eye.}}
  \label{fgr:dinf}  
\end{figure}

\QQ{At long times, diffusion is reestablished although with a significantly smaller diffusion coefficient $D_\infty(\varepsilon)$, which is estimated by the value at $t=80\,t_{\text{B}}$, i.e.~$D_\infty \approx D(80t_{\text{B}})$ (Fig.~\ref{fgr:dinf}). The diffusion coefficient at long times, $D_\infty(\varepsilon)$, has been linked to the free diffusion coefficient $D_0$ \cite{Dean1997,Dean2004,Touya2007,Dean2007}:
\begin{eqnarray}
   \frac{D_{\infty}(\varepsilon)}{D_0} = \mathrm{e}^{- \frac{1}{2} \left (\frac{\mathrm{\varepsilon}}{k_{\text{B}}T} \right)^2}
   \label{eq:DD}
\end{eqnarray}
\QQQ{The most dominant feature of this equation is the dependence on $-(\varepsilon/k_{\text{B}}T)^2$ which is just the ratio of the equilibrium energy of a Gaussian distribution $-\varepsilon^2/k_{\text{B}}T$ and $k_{\text{B}}T$. This first term dominates the temperature-dependence of the barrier, because the typical energies to be crossed for transitions between different regions are essentially temperature-independent, as suggested by a percolation picture (cf. \cite{Dyre1995}).} The simulation findings and theoretical prediction show very good agreement at small $\varepsilon$ and deviations at large $\varepsilon \gtrsim 2 k_{\text{B}}T$ (Fig.~\ref{fgr:dinf}). These deviations are due to the increasingly longer times required to reach the asymptotic long-time value $D_\infty$ which, for $\varepsilon \gtrsim 2 k_{\text{B}}T$, is beyond the simulation time $T_{\text{sim}}$ (Fig.~\ref{fgr:exp_sim}). This is illustrated by the approach of $D(t,\varepsilon)$ toward $D_\infty(\varepsilon)$ for different $\varepsilon$, which is particularly slow and eventually beyond the simulation time $T_{\text{sim}}$ for large $\varepsilon$ (Fig.~\ref{fgr:dinf}, inset).} Note that the simulation time was matched to the experimental recording time, $T_{\text{sim}} \approx T_{\text{exp}}$, in order to obtain equivalent averaging. If the simulation time is increased by an order of magnitude, $T_{\text{sim}} \approx 10\,T_{\text{exp}}$, \QQQ{a significantly better} agreement with the theoretical prediction is observed (Fig.~\ref{fgr:dinf}).

\QQ{At intermediate times, the dynamics are dominated by the slow transition from the initial to the long-time diffusion. This transition can be characterized by the times discussed above: $t_\mu$ and $t_\alpha$ at which the minimum of $\mu(t)$ and the maximum of $\alpha_2(t)$ occur, respectively. These times have been extracted from the simulation data and quantitatively agree with the experimental results (Fig.~\ref{fgr:tmax}).} \QQQ{Based on the Stokes-Einstein equation, the $\alpha$-relaxation time is expected to be inversely proportional to the long-time diffusion coefficient $D_\infty$ \cite{Schweizer2007}. Furthermore, the maximum of the non-Gaussian parameter, i.e.~$t_\alpha$, is typically close to the $\alpha$-relaxation time.} Together with Eq.~\ref{eq:DD}, this suggests $\ln{t_\alpha} \sim (\varepsilon/k_{\text{B}}T)^2$. This is indeed observed (Fig.~\ref{fgr:tmax}). The range of $\varepsilon$ is, however, too small to unambiguously confirm this relation.

\QQ{We now quantitatively compare our experimental and simulation results. This requires to determine the relationship between $\varepsilon$ and the experimentally applied laser power $P$} \Q{as well as the friction coefficient of the particles, $\xi_0^\ast$, which implicitly also depends on the laser power $P$. Due to hydrodynamic effects, the friction coefficient varies with the particles' distance from the water--glass interface \cite{Pagac1996, Leach2009, Sharma2010}. The distance is controlled by a balance between the repulsive particle--wall interaction \cite{Ducker1991,Flicker1993, Walz1997,Gruenberg2001} and the radiation pressure (and gravity) \cite{Ashkin1997, Molloy2002,Bowman2013}, which pushes the particles toward the glass slide and depends on $P$ \cite{Zunke2013}. Both, $\varepsilon(P)$ and $\xi_0^\ast(P)$, are together determined in an iterative procedure which is based on a comparison of the experimental and simulation results and is described in the following.}

\QQ{Although the degree of roughness $\varepsilon$ of the optically-generated potential $U(x,y)$ can be tuned via the laser power $P$ and we expect a linear relationship $\varepsilon \sim P$, $\varepsilon(P)$ cannot easily be determined experimentally.} \Q{Therefore, in a first step, this relation has been estimated using} \QQ{$D_\infty$, which depends on $\varepsilon$ (in the simulations, Figs.~\ref{fgr:exp_sim}, \ref{fgr:dinf}) and $P$ (in the experiments, Fig.~\ref{fgr:DD0}). Since the asymptotic limit $D_\infty$ is not accessible, we use $D_\infty/D_0 \approx D(80t_{\text{B}})/D_0$ for the simulation results and $D_\infty/D_0 \approx D(80t_{\text{B}})/D(0.2t_{\text{B}})$ for the experimental results since the short time limit of the diffusion coefficient is not accessible experimentally (and affected by radiation pressure as described below). An interpolation of $D_\infty(\varepsilon)/D_0$ determined in simulations (Fig.~\ref{fgr:dinf}, red line) was used to assign an $\varepsilon$ to the $D_\infty(P)/D(0.2t_{\text{B}})$ from experiments with different $P$.} \Q{This yields a first approximation for $\varepsilon(P)$.}

The friction coefficient of the particles implicitly also depends on the laser power $P$. A finite $P>0$ will lead to radiation pressure pushing the particles closer to the water--glass interface and hence increases the friction coefficient $\xi_0^\ast > \xi_0$ and reduces the diffusion coefficient $D_0^\ast < D_0$. At short times, the diffusion coefficient tends to a value, $D_{\mathrm{s}}$, which can be used to guide the correction. Although the short-time dynamics are hardly affected by the random potential, the averaging over waiting times $t_0$ (Eq.~\ref{eq:D}) affects $D_{\mathrm{s}}$ \cite{Hanes2012a}, as mentioned above. Thus, the experimental value $D_{\mathrm{s}}=D(t_{\text{s}})$, where $t_{\text{s}} \approx 0.2\,t_{\text{B}}$, was fitted to the corresponding simulation value, which is equally affected by the averaging. The choice of $D_{\mathrm{s}}$ affects $t_{\text{B}}$ and in turn $t_{\text{s}}$ and hence $D(t_{\text{s}})$. \Q{Therefore, the procedure was iterated until consistent relations were obtained.}

This procedure yielded a relation $\varepsilon(P)$ (Fig.~\ref{fgr:eps_scale}), which appears linear up to large $P$ where $\varepsilon$ starts to saturate. The slope is consistent with a previous calibration of a one-dimensional random potential when taking the different illuminated areas into account \cite{Hanes2012}. Furthermore, the iterative procedure provides the friction coefficient; $\xi_0^\ast \approx 1.4\, \xi_0$ for $P>0$ \QQ{and $\xi_0^\ast = \xi_0$ for $P=0$ and the simulations (Fig~\ref{fgr:exp_sim}, inset)}. This implies a scaling factor $f=\xi_0^\ast/ \xi_0$, leading to an effective diffusion coefficient $D_0^\ast = D_0/f$ and an effective Brownian time $t_{\text{B}}^\ast=f t_{\text{B}}$ in the experiments with $P>0$, while in the simulations and experiments with $P=0$, $D_0^\ast=D_0$ and $t_{\text{B}}^\ast=t_{\text{B}}$. Also other procedures have been followed to determine $\varepsilon(P)$ and $\xi_0^\ast(P)$; they all resulted in a linear relation $\varepsilon(P) \sim P$ with slopes within $20\%$ and also very similar $\xi_0^\ast(P)$.

Having determined $\varepsilon(P)$ and corrected the experimental data for radiation pressure effects, we can compare the experimental and simulation results (Fig.~\ref{fgr:exp_sim}). \QQ{While the dynamics at short and long times have been exploited to obtain $\varepsilon(P)$ and $\xi_0^\ast$, a comparison of the intermediate subdiffusive behaviour with the transition from short to long-time diffusion and the corresponding time scales is meaningful. The dynamics at intermediate times indeed quantitatively agree. In addition, the quantitative agreements of the time scales, $t_\mu$ and $t_\alpha$, determined from the experimental and simulation data (Fig.~\ref{fgr:tmax}) have already been discussed.}

\subsection{Comparison to particle dynamics in one-dimensional random and periodic potentials}

\begin{figure}[t!]
  	\centering
   \includegraphics[height=6cm]{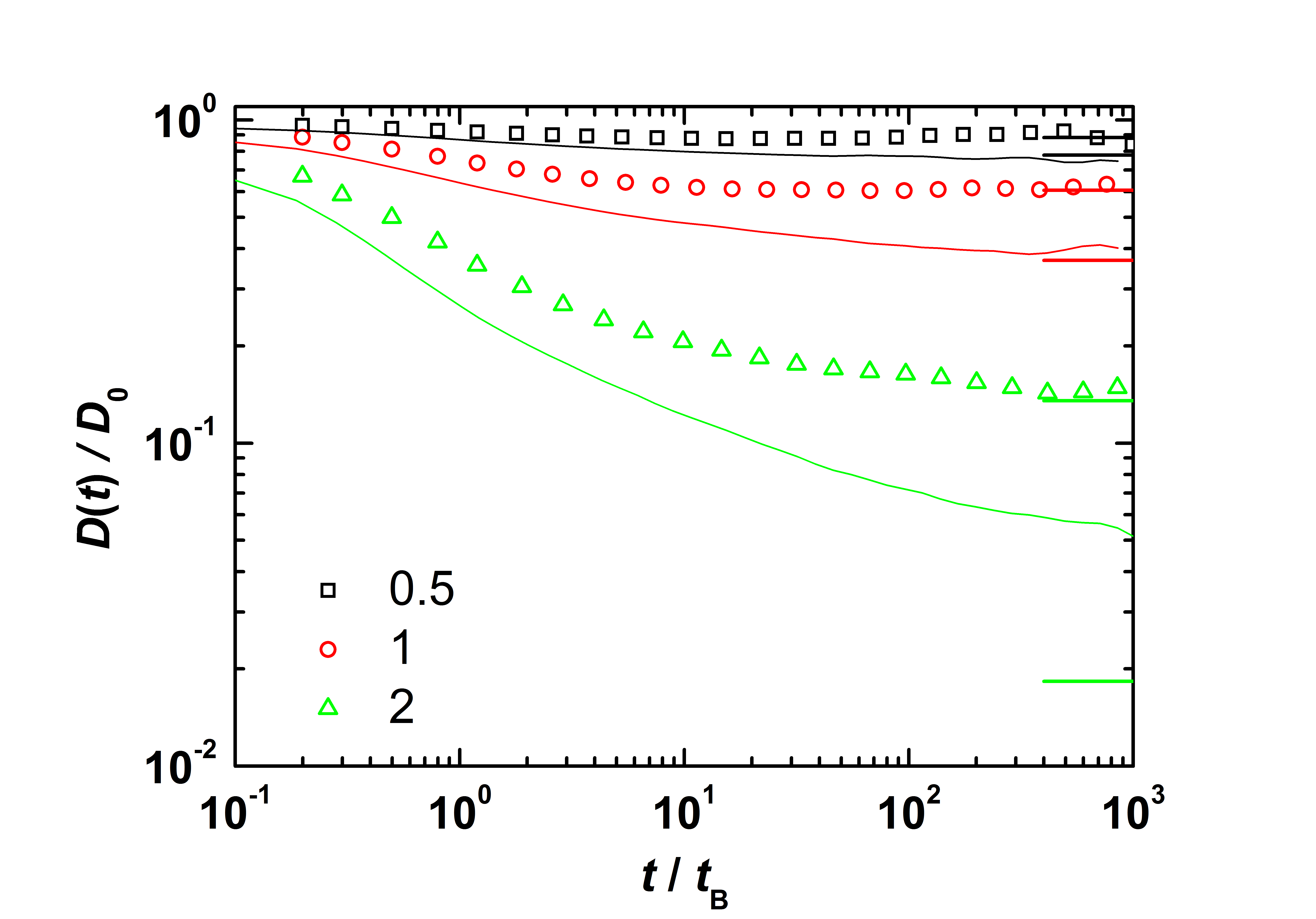}
  \caption{Normalized diffusion coefficient $D(t)/D_0$ as a function of normalized delay time $t/t_{\text{B}}$ for particles in a one-dimensional (lines) and two-dimensional (symbols) random potential with different standard deviations $\varepsilon$ (as indicated) as observed in simulations. \QQ{Solid horizontal lines at large $t/t_{\text{B}}$ correspond to theoretical predictions \cite{Zwanzig1988,Dean2007}.}} 
  \label{fgr:1d2d}
\end{figure}

As in two-dimensional random potentials, in one-dimensional random potentials the particle dynamics also show three distinct regimes: diffusion at short and long times and subdiffusion at intermediate times (Fig.~\ref{fgr:1d2d}) \cite{Hanes2012,Hanes2012a}. The dynamics are much slower in the one-dimensional case. In particular, it takes a much longer time to approach the asymptotic long-time limit. In general, the characteristic times, for example $t_\mu$ and $t_\alpha$, are considerably longer and show a stronger dependence on $\varepsilon$. \QQ{Furthermore, the long-time diffusion coefficient $D_\infty$ is smaller (Fig.~\ref{fgr:dinf}, blue dotted line) \cite{Zwanzig1988}:
\begin{eqnarray}
   \frac{D_{\infty}(\varepsilon)}{D_0} = \mathrm{e}^{- \left (\frac{\mathrm{\varepsilon}}{k_{\text{B}}T} \right)^2}
   \label{eq:DD2}
\end{eqnarray}
In two dimensions, $D_\infty$ is larger because large barriers can be avoided,} \Q{but the exponential dependence on $(\varepsilon/ k_{\text{B}}T)^2$ remains, consistent with the percolation argument.} 

In addition to random potentials, colloidal particles have also been investigated in periodic potentials \cite{Loudiyi1992,Wei1998, Bechinger2000, Dalle-Ferrier2011, Emary2012, Lichtner2012}. In a sinusoidal potential \cite{Dalle-Ferrier2011}, only one barrier height exists and thus the distribution of escape times is narrower. The dynamics at intermediate times exhibit a smaller slope at the inflection point of the mean squared displacement, corresponding to a more pronounced subdiffusive behaviour with a deeper minimum of the exponent $\mu(t)$. On the other hand, long-time diffusion is established earlier as very deep minima are absent.

\section{Conclusions}

We investigated the dynamics of individual colloidal particles in two-dimensional random potential energy landscapes, whose values follow a Gaussian distribution with a standard deviation $\varepsilon$, which characterizes the degree of roughness of the potential. In the experiments, the potential was created using an optical set-up and the roughness $\varepsilon$ was controlled via the laser power $P$. The experimentally observed dynamics agree with our Monte Carlo simulation results. Three distinct regimes have been observed. At short times, the particles exhibit diffusive behaviour within their local minima, in which they remain until they cross a barrier, i.e.~a saddle point, to a neighbouring minima. The wide distribution of barrier heights leads to a significant spread in residence times. In the mean squared displacement this is reflected as a broad subdiffusive region with a relatively large slope at the inflection point at intermediate times. At long times, the hopping between minima resembles a random walk and diffusive dynamics are recovered although with a significantly reduced diffusion coefficient. The long-time diffusion coefficient decreases with increasing degree of roughness $\varepsilon$ in agreement with theoretical predictions \cite{Dean2007}. This decrease is less pronounced than in one-dimensional potential energy landscapes \cite{Zwanzig1988}. This is attributed to the possibility to bypass large barriers in two-dimensions.

The system presented here can also serve as a well-controlled, tunable and easily observable model for other systems, which either explore space or configuration space, i.e.~a potential energy landscape. These systems include crowded systems, such as concentrated colloidal suspensions, supercooled liquids, glasses \cite{Heuer2008, Debenedetti2001, Lubchenko2007,Angell1995, Poon2002, Megen1998, Heuer2005}, or living cells \cite{Weiss2004,Tolic2004,Banks2005}, but also complex potential energy landscapes, such as those suggested in protein folding \cite{Best2011, Dobson1998, Bryngelson1995, Dill1997, Winter2007}.

\section*{Acknowledgement}
We thank J\"{u}rgen Horbach (University D\"{u}sseldorf), Anand Yethiraj (Memorial University, Newfoundland), \QQ{Wilson C.~K.~Poon (The University of Edinburgh) and David Dean (University Bordeaux)} for very helpful discussions and comments. Financial support from the International Helmholtz Research School of Biophysics and Soft Matter (IHRS BioSoft) and the German Science Foundation (DFG) through the German-Dutch Collaborative Research Centre SFB-TR6 (Project Section C7) and the Research Unit FOR1394 (Projects P2, P4) is gratefully acknowledged.

\bibliography{bib}

\end{document}